\documentclass[]{aa}
\usepackage{psfig}

\def\ros{{\sl ROSAT }}
\def\etal{{et\,al.}}

\def\msun{M$_{\odot}$}

\def\mdot{$\dot M$}
\def\grad{$^\circ$}
\def\degs{\ifmmode ^{\circ}\else$^{\circ}$\fi}
\def\amin{\ifmmode ^{\prime}\else$^{\prime}$\fi}
\def\asec{\ifmmode ^{\prime\prime}\else$^{\prime\prime}$\fi}
\def\fss{\hbox{$.\!\!^{\rm s}$}}        
\def\farcs{\hbox{$.\!\!^{\prime\prime}$}}  
\def\h{$^{\rm h}$}
\def\m{$^{\rm m}$}

\newbox\grsign \setbox\grsign=\hbox{$>$}
\newdimen\grdimen \grdimen=\ht\grsign
\newbox\laxbox \newbox\gaxbox
\setbox\gaxbox=\hbox{\raise.5ex\hbox{$>$}\llap
     {\lower.5ex\hbox{$\sim$}}}\ht1=\grdimen\dp1=0pt
\setbox\laxbox=\hbox{\raise.5ex\hbox{$<$}\llap
     {\lower.5ex\hbox{$\sim$}}}\ht2=\grdimen\dp2=0pt
\def\gax{$\mathrel{\copy\gaxbox}$}

\def\rxj{RX\,J1420.4+5334}

\begin{document}
 
   \thesaurus{03          
              (02.01.2;  
               11.01.2;  
               11.09.1;  
               11.14.1;  
               13.25.2)  
             }

   \title{RX J1420.4+5334 -- another tidal disruption event?
         \thanks{Partly based on observations collected at the German-Spanish
          Astronomical Centre, Calar Alto, operated by the MPI f\"{u}r
          Astronomie, Heidelberg, jointly with the Spanish National
          Commission for Astronomy, and the WIYN telescope, operated
          the the University of Wisconsin, Indiana University, 
         Yale University, and the National Optical Astronomy Observatories.}}

   \author{J. Greiner\inst{1}, R. Schwarz\inst{1}, S. Zharikov\inst{2,3},
     M. Orio\inst{4,5}}

   \offprints{J. Greiner, jgreiner@aip.de}
 
  \institute{Astrophysical Institute Potsdam, An der Sternwarte 16,
              14482 Potsdam, Germany
             \and
         Special Astrophysical Observatory, 357147 Nizhnij Arkhyz, Russia
             \and
         OAN, Instituto de Astronom\'{\i}a, UNAM, 22860 Ensenada, M\'{e}xico
            \and
         Osservatorio Astronomico di Torino, Strada Osservatorio 20,
             I-10125 Pino Torinese (TO), Italy
             \and
         Department of Physics, University of Wisconsin, 1150 Univ. Av.,
             Madison, WI 53706, USA
       }

   \date{Received 18 August 2000; accepted 25 September 2000}
 
   \markboth{Greiner et al.}{\rxj\ -- another tidal disruption event?}

   \maketitle

   \begin{abstract}
We have discovered a transient X-ray source, \rxj, which displays a ROSAT 
flux variation of \gax 150 between the \ros All-Sky-Survey in 1990
and a preceding pointed ROSAT observations in July 1990. Optical observations 
suggest a non-active galaxy as the only visible counterpart. We therefore
tentatively identify \rxj\ as a tidal disruption event in a
non-active galaxy.

    \keywords{X-rays: galaxies -- Accretion -- Galaxies: nuclei --
            Galaxies: individual: \rxj
               }

   \end{abstract}
 
\section{Introduction}

Tidal disruption of stars in the gravitational potential of massive black holes
in the centers of galaxies has been recognized long ago as a way to
prove the existence of massive black holes in otherwise non-active galaxies.
During the tidal disruption about 50\% of the mass
of the star are stripped off and get unbound from the star 
(mostly   independent of impact parameter, masses and velocities; 
Ayal \etal\ 2000).
This material can eventually be accreted by the black hole, leading to
a flare of optical to X-ray emission lasting a few months to years.
While originally proposed to explain the activity in AGN (Hills 1975),
a first calculation of the frequency, temporal behaviour and 
spectral signatures of tidal disruption events (Frank \& Rees 1976, 
Young \etal\ 1977, Kato \& Hoshi 1978, Lidskii \& Ozernoi 1979, 
Gurzadyan \& Ozernoi 1980, Gurzadyan \& Ozernoi 1981)
revealed both, the potential of such events to learn more about the inner
regions of galaxies as well as the problems in observing these rare events
(Rees 1988).

Over the recent few years, tidal disruption has attracted renewed attention,
both theoretically and observationally. Loeb \& Ulmer (1997) have studied the
case that a surrounding gaseous envelope could reprocess most of the
radiation from the accretion process into a thermal spectrum of 10$^4$ K,
thus being easily detectable in optical surveys rather than in X-ray surveys.
Magorrian \& Tremaine (1999) have recalculated the stellar disruption rates
in detailed dynamical models of real galaxies, taking into account the 
refilling of the loss cone of stars on disruptable orbits by two-body 
relaxation and tidal forces in non-spherical galaxies. Ayal \etal\ (2000)
have performed simulations of the long-term evolution of tidally disrupted
stars and find that only about 10\% of the stellar mass are actually
accreted by the central black hole.

On the observational side, the discovery of time-variable UV or X-ray emission
in distant, otherwise non-active
galaxies may provide the first indirect evidence of tidal disruption events.
Besides the early report on a UV flare in NGC 4552 (Renzini \etal\ 1995), 
a few very good candidates
have been detected with ROSAT during the recent years, such as 
the $\sim2\times10^{43}$ erg/s outburst in  NGC 5905 
(Bade \etal\ 1996, Komossa \& Bade 1999),
the $>9\times10^{43}$ erg/s flare in RXJ\,1242.6--1119 
(Komossa \& Greiner 1999),
the $\sim10^{44}$ erg/s flare in RXJ\,1624.9+7554 (Grupe et al. 1999),
or the $\sim10^{44}$ erg/s flare in RXJ\,1331.9--3243 (Reiprich \& Greiner 
2000).
Interestingly, in all these ROSAT discoveries the X-ray spectrum of the
flares is extremely soft, of the order of 50-100 eV effective temperature.
The time scales of these events, though not in all cases well constrained, are
of the order of months to years. All these properties, together with the fact 
that none of the corresponding galaxies shows any optical sign of activity
have led to the tentative identification as being
associated with disruption events.

Here we report on the discovery of another transient X-ray source
and the optical and X-ray follow-up observations.

  \begin{figure}
      \vbox{\psfig{figure=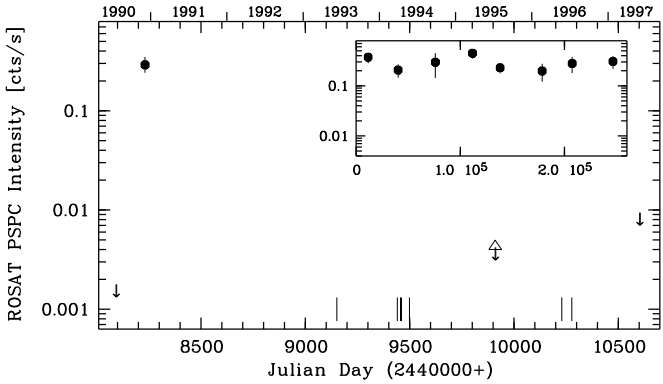,width=\columnwidth,%
          bbllx=1.4cm,bblly=2.7cm,bburx=18.5cm,bbury=12.5cm,clip=}}\par
      \caption[xlc]{Long-term X-ray light curve of \rxj\ with the only
         secure detection during the ROSAT All-Sky-Survey. Because of the
         very soft spectrum we have used a factor 7.8 for the HRI to PSPC 
         count rate conversion (Greiner \etal\ 1997). Upper limits are shown
         with arrows, and the 1995 HRI detection of three photons is marked
         with an open triangle.
         The inset shows the light curve during the three days of the
         All-Sky-Survey.
         The large, vertical dashes at the bottom denote the times of optical 
         observations.
         }
      \label{xlc}
      \vspace{-0.25cm}
\end{figure}

\section{Observational results}

\subsection{ROSAT All-Sky-Survey}

The search for supersoft X-ray sources in the ROSAT PSPC All-Sky-Survey 
(Greiner 1996) revealed a few sources for which neither confirmation
through a ROSAT 
follow-up observation nor a reasonable optical counterpart could be found.
One of these sources is \rxj\ $\equiv$ 1RXS J142024.4+533403 $\equiv$ RBS 1376
(Schwope \etal\ 2000).
It was detected during the PSPC All-Sky-Survey between December 5--8, 1990
at a count rate of  0.28 cts/s, resulting in the collection of a total of
255 counts. There is no significant short-term variability within the
3 day scanning coverage.
The source position has been determined using only photons above 0.25 keV
(to avoid position deterioration by ghost images) to be:
RA (2000.0) = 14\h20\m24\fss2, Decl.
(2000.0) = +53\grad34\amin11\asec\ with an error of $\pm$20\asec.

\begin{figure}
      \vbox{\psfig{figure=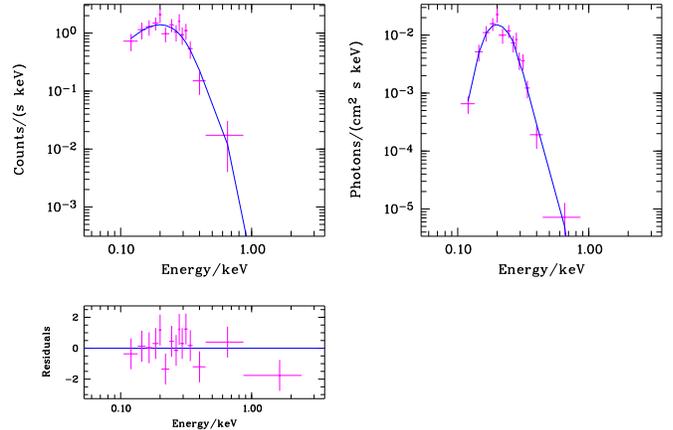,width=\columnwidth,angle=270,%
          bbllx=2.9cm,bblly=1.50cm,bburx=19.5cm,bbury=26.7cm,clip=}}\par
      \caption[]{Fit of a black body model to the X-ray spectrum of \rxj\ 
         as measured during the ROSAT All-Sky-Survey. The lower left panel 
         shows the deviation between model and data in units of $\chi^2$ 
         per spectral bin.
         }
      \label{xspec}
\end{figure}

The spectrum is extremely soft as indicated by the hardness
ratio: $H\!R1=-0.92\pm0.03$ which is defined as 
(H--S)/(H+S), with H (S) being the counts above (below) 0.4 keV over
the full PSPC range of 0.1--2.4 keV.
Source photons were extracted with a radius of
4\amin. The background was chosen at the same ecliptic longitude at
$\approx$1\grad\ distance, corresponding to background photons collected
typically 15 sec before or after the time of the source photons. Standard
corrections were applied using the dedicated EXSAS software package
(Zimmermann \etal\ 1994).
The spectrum fitted with a blackbody model is shown in Fig. \ref{xspec},
and Tab. \ref{xpar} summarizes the fit parameters, also including fits 
with a  power law and an accretion disk model.

  \begin{figure}
      \vbox{\psfig{figure=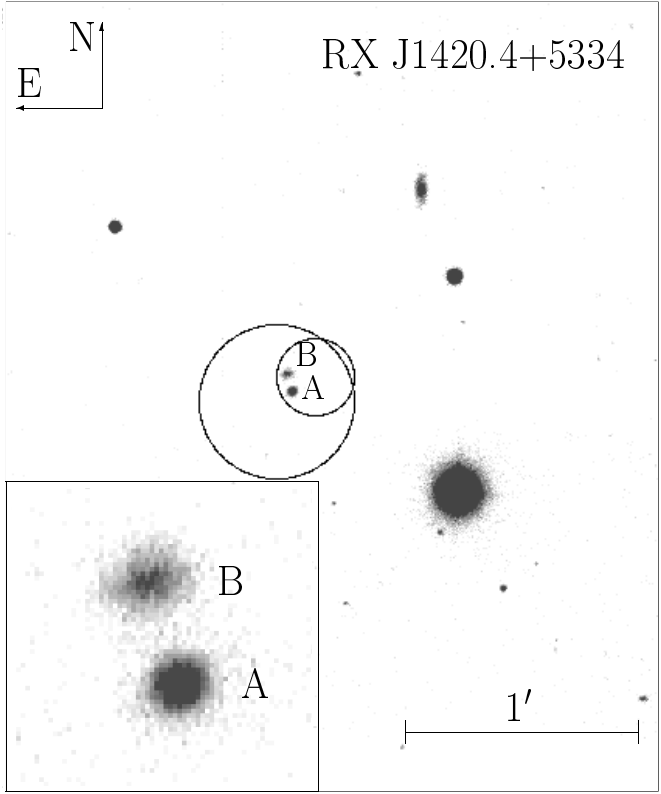,width=8.8cm,%
          bbllx=4.7cm,bblly=9.8cm,bburx=16.2cm,bbury=23.9cm,clip=}}\par
      \caption[]{Finding chart of \rxj\ based on the sum of four B band images
         (total exposure of 2000 sec) taken on May 26, 1996 at the 3.5m WIYN 
         telescope. 
         The large circle marks the 3$\sigma$ X-ray error box of the 
         ROSAT All-Sky-Survey detection (20\asec\ radius), while the smaller 
         corresponds to the position of the photons detected in the HRI
         (10\asec\ radius). The inset shows a zoomed portion of the central
         part.
         }
      \label{fc}
\end{figure}

\subsection{ROSAT pointed observations}

Follow-up pointed ROSAT observations were performed on July 10/11, 1995
and again on June 3, 1997. In both cases the HRI was used. In the 1995
observation we detect three X-ray photons at the position of \rxj,
corresponding to a likelihood of 6.2 (2.5 $\sigma$). Because of this marginal
significance, we do not claim a detection, but rather use 
these detected photons as an upper limit (Fig. \ref{xlc}) on the count rate: 
5.5$\times$10$^{-4}$ HRI cts/s. The mean position of these photons,
RA (2000.0) = 14\h20\m23\fss6, Decl.
(2000.0) = +53\grad34\amin16\asec\ with an error of $\pm$10\asec\
is shown in Fig. \ref{fc} for comparison purposes.
The 1997 observation, with less than half of the 1995 exposure, results
in an upper limit of 1.2$\times$10$^{-3}$ HRI cts/s.

\rxj\ is serendipituously in the field of view of a ROSAT PSPC pointed
observation performed during the calibration phase, though at a very large 
off\-axis angle of 49\amin. No source was detected (using standard
EXSAS commands), giving a 
3$\sigma$ upper limit of 1.8$\times$10$^{-3}$ PSPC cts/s (Fig. \ref{xlc}).



\subsection{Optical observations}

Investigation of the X-ray source position on the POSS revealed only one
bright optical object (labelled ``A'' in Fig. \ref{fc}),
 at RA (2000.0) = 14\h 20\m 24\fss4,
Decl. (2000) = 53\degr\ 34\amin\ 12\asec.
We obtained a low-resolution spectrum of this R $\approx$ 18.8 mag source 
using the 3.5m telescope at Calar Alto on June 13, 1993.
We used the Cassegrain spectrograph
with a 1024$\times$640 RCA chip (pixel size 15 $\mu$m). 
A grating with 240 \AA /mm was used, allowing the range 4000--8500 \AA \ 
to be observed with a FWHM resolution of 10 \AA.
The spectrum was bias and flat-field corrected  and calibrated in 
wavelength using standard MIDAS reduction packages. BD +25\grad 4655 was used 
for flux calibration.

The optical spectrum is characterized by strong absorption
lines of Na\,I 5175\AA, Mg\,I 5890\AA~and H$\alpha$ (though it is near 
the atmospheric A band), typical of 
an elliptical or early spiral type. Using the Mg\,I, Na\,I and H$\alpha$ 
line we derive a redshift of $z=0.147\pm0.001$.
The Balmer absorption lines
and the strong drop of the flux beyond the Ca\,II H/K break argue 
against a classification as BL Lac object. No AGN-like forbidden 
emission lines, like [OIII]$\lambda$5007, are detected.

In order to search for optical variability, we embarked on a more extensive
photometry programme. Observations were done at three locations, the
Sonneberg Observatory (Germany), the Special Astrophysical Observatory (Russia)
and at the WIYN telescope at Kitt Peak (USA). Tab. \ref{log} summarizes the 
logistic details. Throughout these observations we did not found any 
significant optical variability.

The imaging revealed a second, much fainter ($R \approx 20.6$ mag)
source within the X-ray error
circle (labelled ``B'' in Fig. \ref{fc}). During some of the WIYN 
observations, at very good seeing conditions of 0\farcs9, this source is 
readily seen to be extended (see inset of Fig. \ref{fc}). 
Moreover, the intensity distribution over the galaxy disk is very flat.
Since active galaxies typically have a very peaked intensity profile (dominated
by a bright central source), this flat distribution argues for a
non-active galaxy. The shape and radial profile suggest an elliptical or 
early spiral subtype. Size and luminosity arguments then suggest that its 
distance is about a factor 3--5 larger than that of source ``A''.
A lower limit of z$\sim$0.07 can be placed by assuming 
$M_{\rm V} \sim -16$ mag, corresponding to the faint end of the galaxy
luminosity distribution.

Certainly, a spectrum of this galaxy would be very 
helpful to verify these suggestions.

\begin{table} 
\vspace{-0.1cm}
\caption{Log of Observations}
\vspace{-0.25cm}
\begin{tabular}{lllcr}
      \hline
      \noalign{\smallskip}
      ~~Teles-   & ~~~Date  &  Filter/ & D$^{(2)}$ & T$_{\rm Int}$ \\
   ~~cope$^{(1)}$  &          &  Wavelength  & (hrs) & (sec) \\
      \noalign{\smallskip}
      \hline
      \noalign{\smallskip}
 \multicolumn{5}{c}{\bf X-ray} \\
 R 150046P & 1990 Jul 19--22 & 0.1--2.4 keV & 77.7 & 11\,475 \\
 R survey  & 1990 Dec 5--8   & 0.1--2.4 keV & 77.8 & 910 \\
 R 201999P & 1995 Jul 10/11  & 0.1--2.4 keV & 30.7 & 5560 \\
 R 300577H & 1997 Jun 3      & 0.1--2.4 keV & ~\,0.6 & 2150 \\
 \noalign{\smallskip}
 \multicolumn{5}{c}{\bf Optical} \\
 CA 3.5m   & 1993 Jun 13 & 4000--8500 & & 3600 \\
 SO 0.6 m  & 1994 Mar 30 & white      & 2.5 & 30 \\
 SO 0.6 m  & 1994 Apr 14 & white      & 2.9 & 60 \\
 SO 0.6 m  & 1994 Apr 19 & white      & 3.2 & 30 \\
 SO 0.6 m  & 1994 May 27 & white      & 3.0 & 60 \\
 $\!\!$WIYN 3.5m & 1996 May 25 & R/B & 1.0 & 270 \\
 $\!\!$WIYN 3.5m & 1996 May 26 & B & 0.3 & 500 \\
 $\!\!$WIYN 3.5m & 1996 May 27 & B & 0.1 & 500 \\
 SAO 1.0m  & 1996 Jul 14 & V   & 1.5 & 300 \\
 \noalign{\smallskip}
 \hline
 \noalign{\smallskip}
 \end{tabular}

\noindent{
  $^{(1)}$ The abbreviations mean:
  R = ROSAT,
  SO = Sonneberg Observatory (Germany),
  CA = Calar Alto (Spain),
  SAO = Special Astrophysical Observatory (Russia). \\
 $^{(2)}$ Duration of the coverage.
  }
   \label{log}
  \vspace{-0.2cm}
 \end{table}

  \begin{figure*}
    \vbox{\psfig{figure=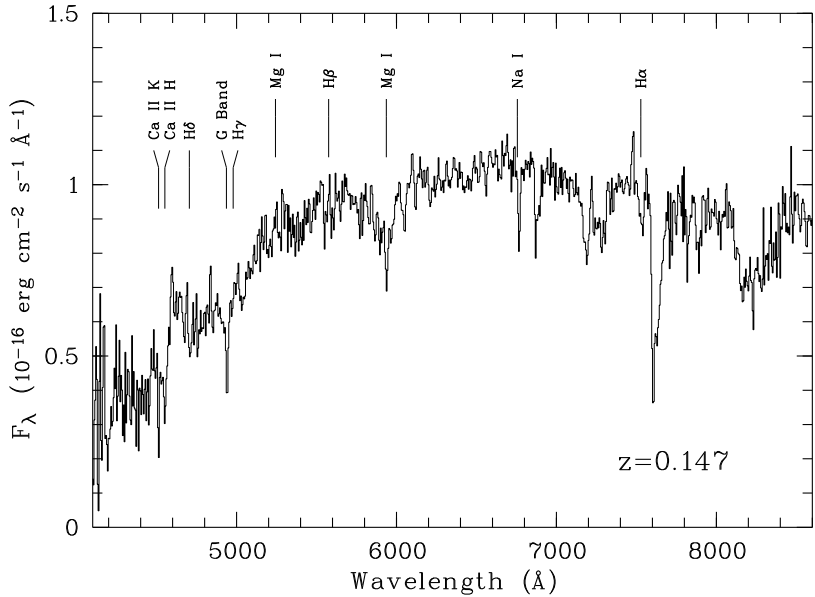,width=12.cm,%
           angle=270}}\par
     \vspace*{-1.4cm}\hfill
     \parbox[b]{60mm}{
     \caption[osp]{Optical spectrum of \rxj, source ``A''. Prominent absorption
    lines are marked.
         }}
      \label{ospec}
\end{figure*}

   \begin{table*}
     \caption{Spectral fit results of \rxj}
      \vspace{-0.2cm}
      \begin{tabular}{llccc}
      \hline
      \noalign{\smallskip}
      Model$^{(1)}$ & \hspace{2.cm} Fit parameters$^{(2)}$ & $\chi^2_{red}$ & 
             Flux$^{(3)}$ (0.1-2.4 keV) & Flux$^{(3)}$ (bol.) \\
       \noalign{\smallskip}
      \hline
      \noalign{\smallskip}
      bbdy, free fit & $kT$ = 28$\pm$10 ~~~ $N_{\rm H}$ = 2.0$\pm$0.1 & 0.96 & 
        7.9$\times$10$^{-12}$ & 1.6$\times$10$^{-11}$ \\
      bbdy, $N_{\rm H}$ = $N_{\rm H}^{\rm gal}$ & $kT$ = 38$\pm$10 & 0.90 &
        8.8$\times$10$^{-13}$ & 4.1$\times$10$^{-12}$  \\
      powl, free fit & $\alpha = -7.5\pm$1.5 ~~~ $N_{\rm H}$ = 2.8$\pm$1.0 & 
        0.96  & 2.4$\times$10$^{-12}$  & -- \\
      powl, $N_{\rm H}$ = $N_{\rm H}^{\rm gal}$ & $\alpha = -4.8\pm$0.5 & 1.07
          &  8.9$\times$10$^{-13}$ & -- \\
      dibb, free fit & \mdot = 0.01$\pm$0.005 ~~~ $M$ = 10$^6$ (fix) ~~~ 
            $N_{\rm H}$ = 6.6$\pm$1.5 & 0.94
          & 5.0$\times$10$^{-10}$  & 1.8$\times$10$^{-6}$ \\
      dibb, $N_{\rm H}$ = $N_{\rm H}^{\rm gal}$ & \mdot = 2.6$\pm$0.1 ~~~ 
        $M$ = 10$^6$ (fix)  & 0.90 
          & 9.2$\times$10$^{-13}$ & 6.2$\times$10$^{-12}$ \\
      dibb, $N_{\rm H}$ = $N_{\rm H}^{\rm gal}$  &
          \mdot = 1.8$\pm$0.1 ~~~ $M$ = 7$\times$10$^5$ ~~~ 
              Norm$^{(4)}$=2.9$\times$10$^{-10}$ (fix) & 0.90
          & 9.2$\times$10$^{-13}$ & 6.4$\times$10$^{-12}$ \\
      \noalign{\smallskip}
      \hline
      \noalign{\smallskip}
  \end{tabular}

\noindent{\small 
      $^{(1)}$ The galactic column density is $N_{\rm H}^{\rm gal}$ = 
          1.2$\times 10^{20}$ cm$^{-2}$ (Dickey \& Lockman 1990).
          The model abbreviations have the following meaning: 
          bbdy = black body, powl = power law, dibb = disk black body model.\\
       $^{(2)}$ Units are eV for temperatures, \msun\ for masses, 
            $10^{20}$ cm$^{-2}$ for $N_{\rm H}$. \\
       $^{(3)}$ Observed fluxes in units of erg cm$^{-2}$ s$^{-1}$ 
            (absorption-corrected). \\
       $^{(4)}$ Normalization is chosen such that distance is equal to
          $D_L = c/H * z$ = 588 Mpc (and inclination assumed to be face-on).
       }
   \label{xpar}
   \end{table*}

\section{Discussion}

We have presented evidence for a transient X-ray source with an
amplitude of a factor of $>$150 within 6 months. The only optical 
counterparts down to 23rd magnitude are two galaxies, the brighter of which
is at a redshift of z=0.147, and the fainter one probably a factor 3--5 
more distant.
If the X-ray flare occurred in the brighter
galaxy, the unabsorbed, bolometric X-ray luminosity is 
2.5$\times$10$^{44}$ erg/s (using the disk blackbody model with only
galactic foreground absorption and a normalization which fits the z=0.147
distance; see last line in Tab. \ref{xpar}). It should be mentioned that
this is a lower limit since most probably we did not catch the peak
of the flare, and the X-ray emission may have suffered additional
intrinsic absorption.
If the flare occurred in the fainter galaxy (B) then the luminosity would 
probably be even larger, but certainly not lower than 6$\times$10$^{43}$ erg/s 
(if at z$\sim$0.07).

Though it is impossible to claim with certainty that this evidence proves
a tidal disruption origin, we will in the following, by the principle of 
exclusion, discuss possible alternatives (for a more extensive discussion of
alternatives see Komossa \& Bade 1999).
(1) An active galaxy as the most common origin of large X-ray variability 
  is very unlikely given our optical observations (see above section 2.3).
(2) X-ray afterglows of gamma-ray bursts (GRB) may appear as new, bright 
  X-ray sources on the sky. However, all of the observed afterglows have 
  faded rapidly (with time$^{-0.8...-2.1}$; see Greiner 2000).
  The ROSAT all-sky survey data suggests a rather flat light curve over
  about 3 days, inconsistent with a GRB afterglow.
(3) This flat, 3-day long light curve also excludes neutron star merger
  events which are predicted to reach peak luminosities of 10$^{44}$ erg/s, 
  but should fade by a factor of $\sim$3 within $\sim$2 days 
  (Li \& Paczynski 1998).

Thus, a tidal disruption scenario seems a logical consequence. Moreover, 
it is even plausible: Spectral fitting of a simple disk blackbody model
(Shakura \& Sunyaev 1973) gives a consistent picture concerning X-ray
temperature, mass of the central object ($\sim$10$^6$ \msun), luminosity 
and distance  (see last row of Tab. \ref{xpar}).

\begin{acknowledgements}
We are grateful to E. Pavlenko (Crimean Observatory) and W. Wenzel (Sonneberg)
for initial optical observations in 1994, as well as to N. Borisov for
help in observations conducted at SAO.
We appreciate enlightening discussions with M. Salvato (AIP) on the
radial profiles of galaxies.
JG was partly supported by the German Bundes\-mini\-ste\-rium f\"ur 
Bildung, Wissenschaft, Forschung und Technologie (BMBF/DLR) under contract 
50\,QQ\,96\,02\,3. The \ros project
was supported by 
BMBF/DLR and the Max-Planck-Society.
\end{acknowledgements}


\begin{thebibliography}{}

\bibitem[]{alp00} Ayal S., Livio M., Piran T., 2000, ApJ (in press;
astro-ph/0002499)

\bibitem[]{b96} Bade N., Komossa S., Dahlem M., 1996, A\&A 309, L35


\bibitem{dl90} Dickey J.M., Lockman F.J., 1990, ARAA 28, 215

\bibitem{fr76} Frank J., Rees M.J., 1976, MNRAS 176, 633

\bibitem[]{g96} Greiner J., 1996, in Supersoft X-ray Sources,
ed. J. Greiner, Lect. Notes in Phys. 472, Springer, p. 285

\bibitem[]{gbl97} Greiner J., Bickert K.F., Luthardt R., Viotti R., 
  Altamore A., Gonz\'alez-Riestra R., Stencel R.E., 1997, A\&A 322, 576

\bibitem[]{ghv00} Greiner J., 2000, 
http:/$\!$/www.aip.de/People/JGreiner/grb.html 


\bibitem[]{g99} Grupe D., Thomas H.-C., Leighly K.M., 1999, A\&A 350, L31

\bibitem[]{g80} Gurzadyan V.G., Ozernoi L.M., 1980, A\&A 86, 315

\bibitem[]{g81} Gurzadyan V.G., Ozernoi L.M., 1981, A\&A 95, 39

\bibitem[]{kh78} Kato M.,  Hoshi R.,  1978, Prog. Theor. Phys. 60, 1692

\bibitem[]{kb99} Komossa S., Bade N., 1999, A\&A 343, 775

\bibitem[]{kg99} Komossa S., Greiner J., 1999, A\&A 349, L45

\bibitem[]{lp98} Li L.-X., Paczynski B., 1998, ApJ 507, L59

\bibitem[]{lo79} Lidskii V.V., Ozernoi L.M., 1979, Pis'ma Astron. Zh. 5, 28 
  (Soviet Astron.  Lett. 5, 16)

\bibitem[]{mt99} Magorrian J., Tremaine S., 1999, MNRAS 309, 447

\bibitem[]{r88} Rees M., 1988, Nature 333, 523

\bibitem[]{rg00} Reiprich T.H., Greiner J., 2000, in Black Holes in Binaries 
   and Galactic  Nuclei,
  eds. L. Kaper, E.P.J. van den Heuvel, and P.A. Woudt, Springer (in press)

\bibitem[]{rgs95} 
Renzini\,A.,\,Greggio\,L.,\,Di\,Serego-Alighieri\,S.\,et\,al.,\,1995,\,Nat.\,378,\,39

\bibitem[]{shl00} Schwope A.D., Hasinger G., Lehmann I., \etal\ 2000, 
   AN 321, 1

\bibitem[]{ss73} Shakura N.I., Sunyaev R.A., 1973, A\&A 24, 337

\bibitem[]{ysw77} Young P.J., Shields G.A., Wheeler J.C., 1977, ApJ 212, 367

\bibitem{zbb94} Zimmermann H.U., Becker W., Belloni T., D\"obereiner S., 
 Izzo C., Kahabka P., Schwentker O., 1994, MPE report 257

\end{thebibliography}
\end{document}